\documentclass[aps, prx, floatfix]{revtex4-2}
\usepackage{graphicx} 
\usepackage[hmargin=1.0in, vmargin=1.0in]{geometry}

\newcommand{\sub}[1]{${}_{#1}$}

\begin{document}

\title{Removing Atmospheric Carbon Dioxide Using Large Land Or Ocean Areas Will Change Earth Albedo And Force Climate}

\author{J. B. Marston}
\email{Author to whom correspondence should be addressed: marston@brown.edu}
\affiliation{Brown Theoretical Physics Center and Department of Physics, Brown University, Providence, Rhode Island 02912, USA}

\author{Daniel E. Ibarra}
\affiliation{Department of Earth, Environmental and Planetary Sciences and the Institute at Brown for Environment and Society, Brown University, Providence, Rhode Island 02912, USA}

\begin{abstract}
When large surface areas of the Earth are altered, radiative forcing due to changes in surface reflectance can drive climate change. Yet to achieve the necessary scale to remove the substantial amounts of carbon dioxide from the atmosphere relevant for ameliorating climate change, enhanced rock weathering (ERW) will need to be applied to very large land areas.  Likewise, marine carbon dioxide removal (mCDR) must alter a large fraction of the ocean surface waters to have a significant impact upon climate. We show that surface albedo modification (SAM) associated with ERW or mCDR can easily overwhelm the radiative forcing from the decrease of atmospheric CO\sub{2} over years or even decades.   A change in albedo as small as parts per thousand has a radiative impact comparable to the removal of 10 tons of carbon per hectare.  SAM via ERW can be either cooling or warming. We identify some of the many questions raised by radiative forcing due to these forms of CDR. 
\end{abstract}

\date{\today}

\maketitle

\section{Introduction}
\label{Introduction}
There are two principal approaches to climate intervention, the deliberate manipulation of the Earth's climate system to ameliorate global warming that is also known as geoengineering.  The first type of intervention is to remove long-lived greenhouse gases presently in the atmosphere.  Carbon dioxide removal (CDR) has received the most attention, but consideration has also been given to reducing other gases such as methane.  The other intervention is known as solar radiation management (SRM).  Here the goal is to reduce the amount of solar radiation absorbed by the oceans, land, ice, and/or atmosphere.  Both CDR and SRM are controversial (SRM particularly so).  Recently, the American Geophysical Union released ethical guidelines for both types of climate intervention \cite{AGU.2024}.  Typically, CDR and SRM are considered separately.  Here we argue that some forms of CDR are inextricably linked to a form of SRM known as surface albedo modification (SAM).  

To achieve the scale necessary to remove gigatons of carbon dioxide from the atmosphere relevant for ameliorating climate change, the CDR strategy known as enhanced rock weathering (ERW) will need to be applied to very large land areas.  Likewise marine carbon dioxide removal (mCDR) must alter a large fraction of the ocean surface to have a significant effect.  When large surface areas are modified, significant changes in radiative forcing due to changes in albedo, the fraction of reflected sunlight, must be considered.  Unappreciated until now, we demonstrate the potential that surface albedo modification (SAM) can overwhelm the reduction in radiative forcing from the decrease of atmospheric CO\sub{2} due to ERW over timescales of decades.  Thus, it is essential to consider radiative impacts of these forms of CDR \cite{Taylor.2024}.  Importantly, the sign of SAM via ERW  applications can be either positive or negative depending on the choice of materials added to the soil and other factors.  For instance, depending on the albedo of the untreated soil, mafic or ultramafic minerals could decrease soil albedo and thus contribute to warming.  Conversely, whitish minerals such as wollastonite have high albedo and are expected to increase the reflectance of sunlight and cool the Earth.  A few previous studies have considered the radiative forcing due to carbon dioxide removal (CDR) but via reforestation \cite{Mykleby.2017,Weber.2024,Kristensen.2024} not ERW or mCDR.  This previous work argues that a decrease in albedo due to reforestation can cancel out a substantial portion of the benefits from CO\sub{2} removal by trees.  

There are multiple potential benefits to the combination of CDR with (cooling) SAM: (1) Immediate cooling by increased surface reflectance (albedo) would be followed by (2) the slow removal of atmospheric CO\sub{2} thereby addressing the root cause of climate change. Furthermore, ERW (3) offers co-benefits to agriculture via soil amendments [e.g. \cite{Beerling.2024}] from crushed rock potentially making deployment over large areas attractive to farmers. Finally, (4) ERW and mCDR have already been deployed in field experiments around the world with little or no controversy in contrast with other approaches to solar radiation management (SRM) \cite{Szerszynski.2013,Contzen.2024,Sugiyama.2024,Flavelle.2024}.  

We argue below that the radiative forcing due to SAM could greatly exceed that from CDR by ERW and possibly mCDR as well.  A campaign of field measurements at existing ERW and mCDR sites could systematically measure SAM due to these forms of CDR.  Research questions that we identify, when addressed, will answer a number of open questions about the radiative impacts of ERW and mCDR and the possibility of jointly deploying and optimizing ERW or mCDR and SAM at scale to maximize global cooling.

\section{Carbon Dioxide Removal and Surface Albedo Modification}
\label{CDR+SAM}
Radiative forcing, $\Delta F$, is the change in the downward minus upward radiative flux.  It is the base driver of climate change. Changes in both short-wave (visible light) and long-wave (infrared) radiation must be considered. Any type of ERW or mCDR scaled to the magnitude needed to remove substantial amounts of atmospheric carbon dioxide would at the same time alter large land or ocean surface areas, potentially leading to significant radiative forcing.  These forms of CDR have been classified as ``once-through'' because large amounts of material such as ground rock are used once in contrast to ``cyclic'' CDR processes such as direct air capture \cite{Taylor.2024} that reuse carbon capture materials and consequently are energy intensive but affect smaller land areas.  Here we briefly review ERW and then discuss several types of marine CDR before turning to SAM.  

\subsection{Enhanced Rock Weathering (ERW)}
\label{ERW}
ERW seeks to capture atmospheric CO\sub{2} using minerals that are out of equilibrium at Earth's surface to generate alkalinity via surficial weathering reactions \cite{Rinder2021,Beerling.2020,Beerling.2024}. Practically speaking, this can be done by mining, crushing, and grinding rock containing reactive minerals and spreading them over large areas in agricultural or managed land settings with the possibility of repeat applications on a designated (likely seasonal) timescale \cite{Beerling.2024}. There, the rock dust reacts with atmospheric and soil CO\sub{2}, the latter of which is derived from root and microbial respiration, to form relatively stable carbonates or (in most cases) carbonate species, in the form of bicarbonate in most settings, in solution that is delivered to the ocean via rivers and thus enhances ocean alkalinity \cite{Oh2006, Zhang2024}. Such soil amendments may, in certain settings, also enhance soil organic carbon stocks \cite{Manning2024}. ERW attempts to speed up these natural weathering processes by overcoming the kinetic and physical limitations via enhanced surface area in finely ground rock applied to the top of soils in order to capture significant quantities of atmospheric CO\sub{2} over years to decades \cite{Beerling.2024,Campbell.2022,Breunig.2024,Rinder2021,Goll.2021}.

Various specific rocks and minerals have been proposed for enhanced surficial weathering.   Wollastonite reacts with carbon dioxide exothermically as summarized by the following reaction \cite{Welbel.2023}:
\begin{eqnarray}
{\rm CaSiO}_3 + 2 {\rm CO}_2 + {\rm H}_2{\rm O} &\rightarrow& {\rm Ca}^{2+} + 
2 {\rm HCO}_3^- + {\rm SiO}_2
\nonumber \\
&\rightarrow& {\rm CaCO}_3 + {\rm SiO}_2 + {\rm CO}_2 + {\rm H}_2{\rm O}
\label{wollastonite+CO2}
\end{eqnarray}
Based on Equation \ref{wollastonite+CO2}, the theoretical carbon removal efficiency of wollastonite is $0.76$ T CO\sub{2} per ton of wollastonite applied. In actual experiments the rate of CO\sub{2} sequestration has been shown to vary between $87$ and $255$ T CO\sub{2} per KT of wollastonite for each year of soil amendments \cite{haque2020optimizing,wood2023impacts}. 

Dark mafic and ultramafic rocks such as basalt or peridotite may decrease soil albedo whereas whitish minerals such as wollastonite should increase the reflectance (Figure \ref{minerals}).  Amendment of soils with biochar is another way to sequester carbon \cite{Bai.2019}. Biochar is typically dark in color, may reduce soil albedo and, if deployed over large areas, increase global temperatures along the lines described below in Section \ref{Model}.

\subsection{Marine Carbon Dioxide Removal (mCDR)}
\label{mCDR}
Because the ocean contains about 50 times as much carbon as the atmosphere and absorbs about half of anthropogenic emissions of CO\sub{2}, it is natural to consider how changes in the oceans could enable the absorption of additional carbon from the atmosphere. Proposals to modify
ocean chemistry to transform atmospheric CO\sub{2} into carbonate ions dissolved in surface waters
(ocean alkalinity enhancement) range from the addition of alkaline materials to surface waters to electrochemistry \cite{NASEM.2022, Lebling.2022, Campbell.2022}. The residence times of added minerals in surface waters, and possible induced changes in biology, need to be considered. How such changes in surface waters might alter the albedo is unclear.  Seawater has low albedo as it absorbs most light; thus, even the addition of ultramafic minerals could brighten the sea surface. In contrast to soil amendments, materials added to oceans reflect light even at considerable depths below the surface.

Another approach seeks to fertilize surface waters with nutrients such as iron to stimulate the biological carbon cycle in the hope that some of the carbon absorbed from the atmosphere will make its way to deep waters \cite{NASEM.2022,Buessler.2023}.  Surface blooms of diatoms or algae would increase the albedo resulting in cooling SAM.  Possible increases in ocean albedo due to the release from marine organisms of dimethyl sulfide that can form cloud condensation nuclei should also be studied.

The effects of a mCDR treatment on the surface ocean are expected to dissipate faster than the effects of ERW on land as materials and alkaline-rich waters will be ventilated to depths below the ocean mixed layer.  Given the uncertainties in the radiative impacts of mCDR, we focus on ERW for the remainder of this Perspective.

\subsection{Surface Albedo Modification (SAM)}
\label{SAM}
Most SRM schemes to cool Earth have focused on reflecting sunlight by stratospheric aerosol injection or by brightening marine clouds. More speculatively the thinning of cirrus clouds to enhance long-wave emission has also been considered. These schemes are controversial \cite{Szerszynski.2013,Contzen.2024,Sugiyama.2024,Flavelle.2024}. Surface albedo modification (SAM, also known as land radiative management or LRM \cite{Seneviratne.2018}) has by contrast received relatively little attention \cite{NAP.2015,Monitor.2021,Johnson.2022} despite the fact that the built environment (asphalt, buildings, parks, etc.) already alters microclimates in easily noticeable and measurable ways (e.g., the urban heat island effect) \cite{Ouyang.2022}. For this reason, SAM (in conjunction with ERW or mCDR) could be less controversial than other forms of SRM.

\begin{figure}[ht]
\begin{center}
\includegraphics[width=6cm]{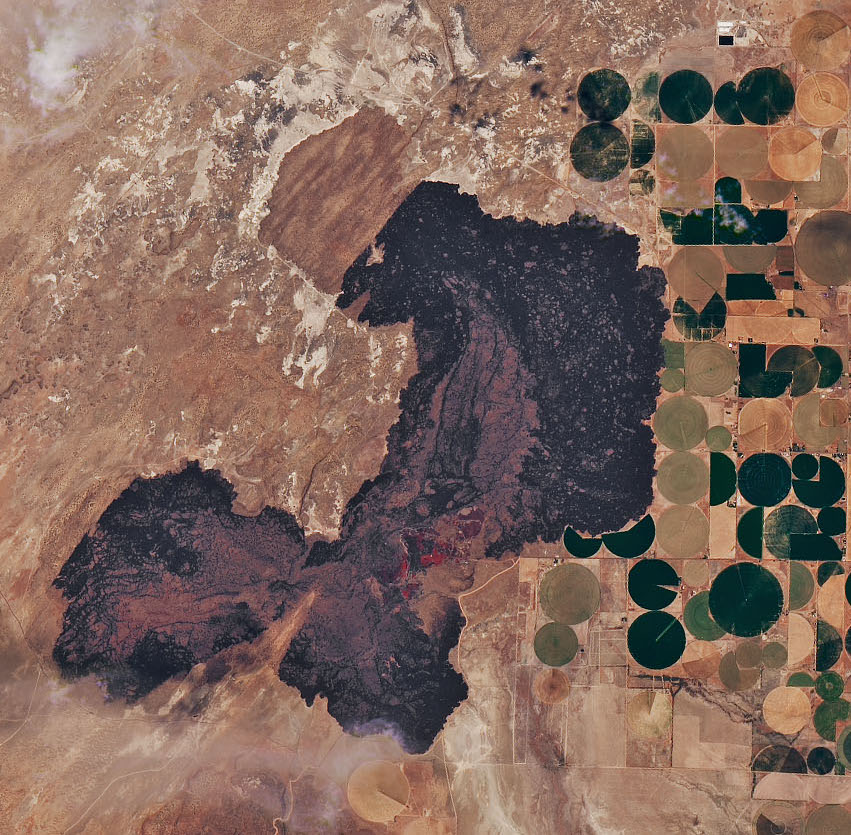}
\hskip 1cm
\includegraphics[width=8cm]{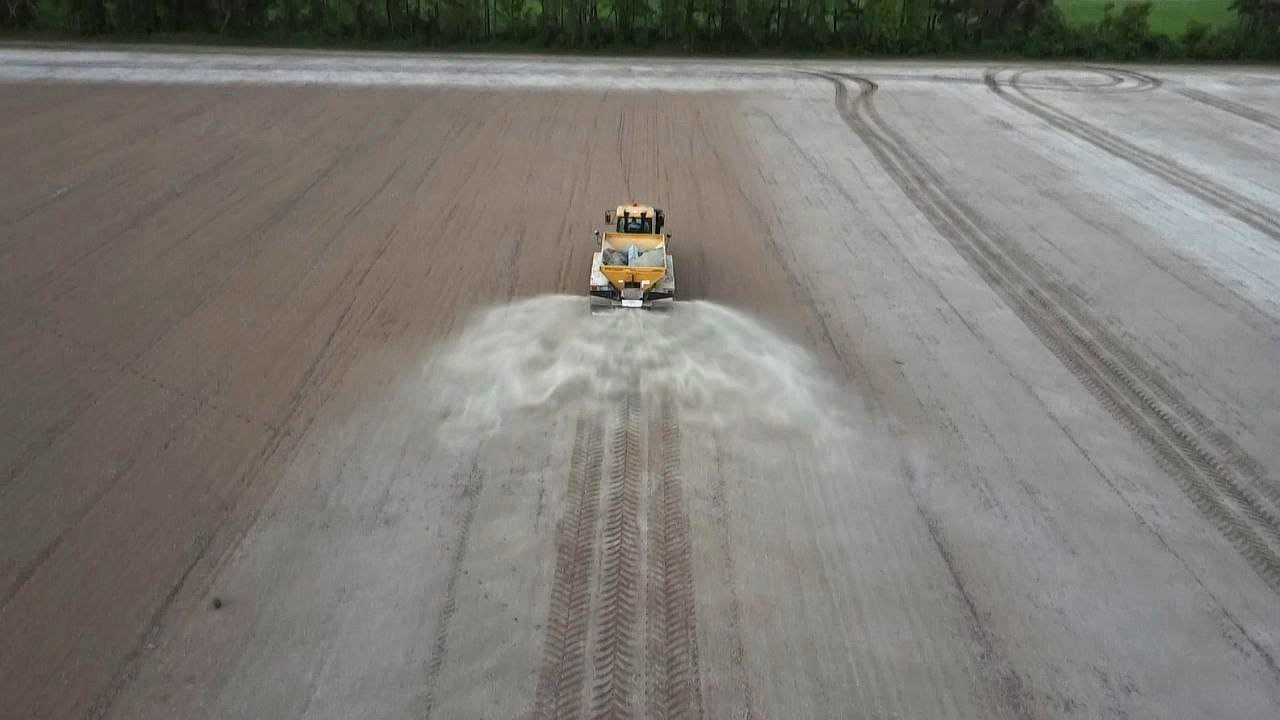}
\end{center}
\caption[x]{Left: View from space on July 12, 2022 of the 45 km$^2$ Ice Springs basalt lava flow in Utah, part of the ``Cinders'' volcanic complex.  The dark, low albedo, mafic basalt flow contrasts with some of the center pivot irrigation areas that are likely planted with winter wheat or lying fallow.  The albedo of the flow increases as it weathers (lighter region near top that is separated from the younger basalt by a diagonal fault line).  (Photo credit: NASA Earth Observatory.)  Right: Crushed wollastonite being spread by a tractor shows an increase in reflectance and hence surface albedo (SAM).  (Photo credit: Canadian Wollastonite.)}
\label{minerals}
\end{figure}

The standard criticism of SAM implementations is that modification of enormous surface areas is required to have a significant impact on the global climate \cite{NAP.2015}. However, as we show below ERW will have potentially strong SAM effects (see Figure \ref{minerals}) that have thus far been overlooked in studies. As large areas are required for gigaton scale carbon removal, a judicious choice of minerals may realize cooling SAM in conjunction with ERW. As a point of reference, an increase of the albedo by 0.1 on the 4.8 gigahectares of land used worldwide for agriculture would result in Earth cooling by roughly $1 ^\circ$C assuming a transient climate sensitivity of $0.7 ^\circ$C / (W/m$^2)$. For ERW, how much rock dust remains on the soil surface is of crucial importance. Whitish alkaline minerals could be deliberately left on the top of soils to maximize cooling SAM. Even a perfectly mixed tilling of a typical deployment of 50 T / ha of crushed rock into the top 15 cm of soil will result in a volume fraction of about 2\% rock (assuming the soil and rock dust have equal densities of $1.5$ gm/cm$^3$). Thus an increase in albedo of order 1\% may be expected from the addition of whitish rock to fully tilled soils.

\section{A Zero-Dimensional Model of Combined CDR and SAM} 
\label{Model}
The radiative forcing due to the modification of land or ocean surfaces can dwarf that due to CO\sub{2} uptake for years after treatment.  As a concrete illustration consider ERW the absorbs CO\sub{2} at the rather optimistic rate \cite{Breunig.2024} of $10$T CO\sub{2}/ha/year, and for the purpose of illustration consider the land area required to remove $1$ GT CO\sub{2}/year, namely $10^6$ km$^2$. (Lower rates of CDR only increase the relative importance of SAM.) Note that the relative importance of SAM versus ERW does not depend on the size of the treated area as both scale linearly with the area of the treated surface.  

Long-wave radiative forcing due to CDR is spread out over the globe because CO\sub{2} is a well-mixed gas, whereas the intense short-wave forcing due to SAM is localized over the treated areas.  In the following, we assume that Earth's energy budget and atmospheric circulation mix this forcing globally.  Much more sophisticated climate model simulations of SAM's regional and global impact on surface temperatures and the water cycle are needed \cite{Cheng.2023,Cheng.2024}.  Nonetheless, a zero-dimensional energy balance model suffices for rough estimates of global effects.  Field measurements can be used to calibrate more sophisticated climate models.  

The assumed treated area of $\Delta A = 10^6$ km$^2$ is a fraction $\Delta A / A = 0.2\%$ of Earth's total surface area of $A = 5.1 \times 10^8$ km$^2$.  The average solar irradiance $G$ is $1/4$ of the solar constant.  At the top of the atmosphere therefore $G = 340$ W/m$^2$.  Ignoring for the sake of simplicity latitude and scattering of light by the atmosphere and clouds (which reduces this number \cite{Masson.2021} to an average of $185$ W/m$^2$), a change in albedo $\Delta a$ alters the radiative forcing $\Delta F_{SAM}$, averaged over the Earth's surface, by 
\begin{eqnarray}
\Delta F_{SAM} = - \Delta a~ G~ \frac{\Delta A}{A} = -(0.7~ {\rm W/m}^2) \times \Delta a\ .
\label{SAM-forcing}
\end{eqnarray}  
This value can be compared to the radiative forcing $\Delta F_{CDR}$ due to the removal of $\Delta C = -1$GT of CO\sub{2}.  There are about $C = 3,500$ GT CO\sub{2} presently in the atmosphere, and upon making a linear approximation to the actual logarithmic dependence \cite{Myhre.1998,Mlynczak.2016} on CO\sub{2} concentration, the change in the radiative forcing due to the reduction in atmospheric CO\sub{2} by 1 GT is approximately  
\begin{eqnarray}
\Delta F_{CDR} = (5.35~ {\rm W/m}^2) \times \frac{\Delta C}{C} = -1.5 \times 10^{-3}~ {\rm W/m}^2\ .
\label{CDR-forcing}
\end{eqnarray} 
Thus, comparing Equation \ref{SAM-forcing} to Equation \ref{CDR-forcing} it is evident that even a tiny brightening of the albedo from, say, $0.300$ to $0.302$, or $\Delta a = 2 \times 10^{-3}$, would change the SAM radiative forcing by an amount comparable to that due to the CO\sub{2} removed by ERW in one year. (A simple model for the removed CO\sub{2} is $\Delta C = -k~ t~ \Delta A$ where $t$ is time and $k$ is a reaction rate, e.g. $k = 10$ T/ha/year, though the actual time dependence will undoubtedly be more complex.)  Note however that the removed CO\sub{2} has a long-term continuing effect whereas how the albedo of the treated soils would change over time is unknown (and is one of the questions we identify below that must be addressed). Repeat application of ground rock, currently proposed and implemented in recent field experiments \cite{Beerling.2024}, would presumably also alter the albedo in a roughly additive manner to a limit or saturation, or at a minimum would alter albedo in a sustaining manner.  
\begin{figure}[h!]
\begin{center}
\includegraphics[width=8cm]{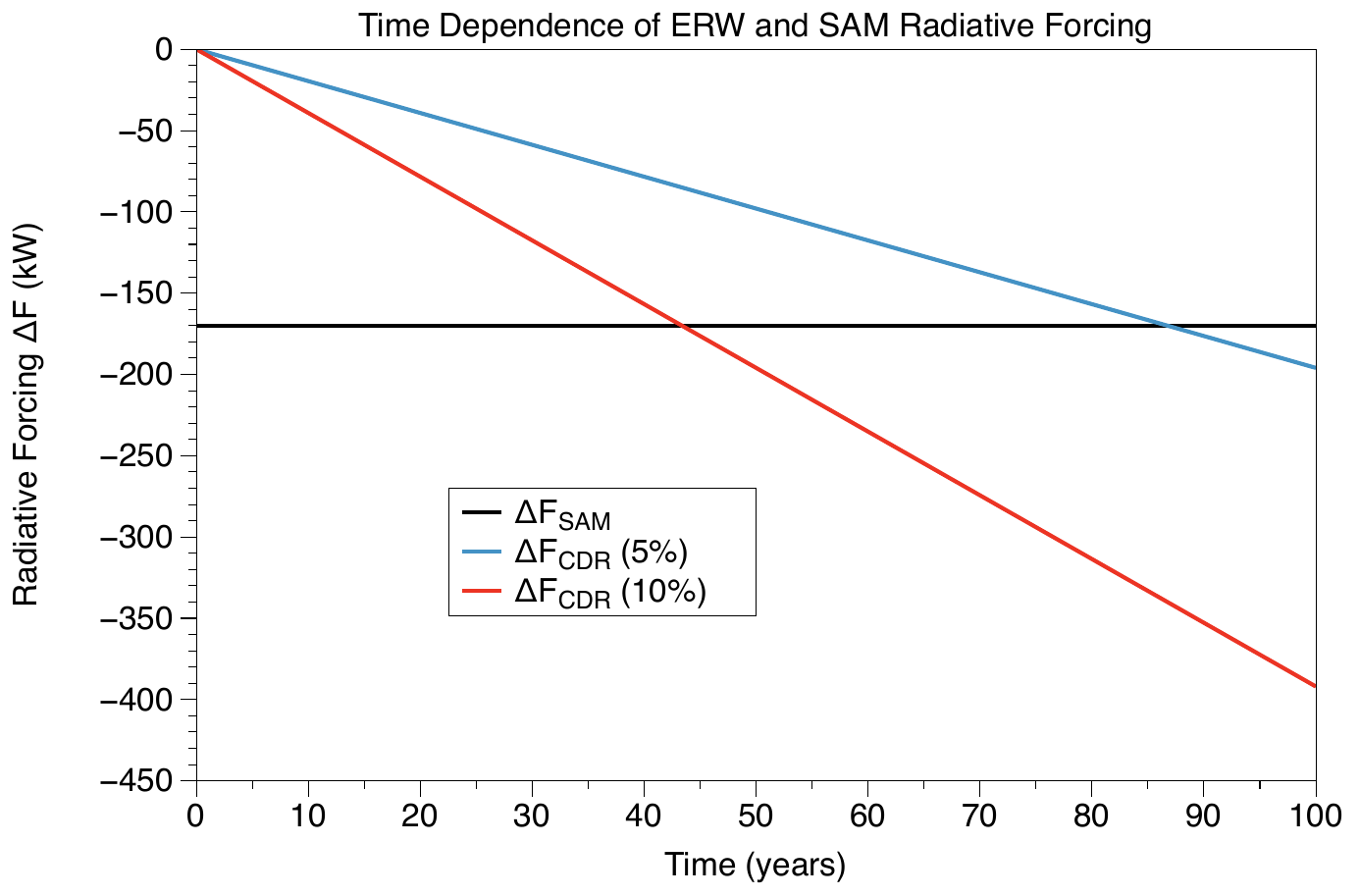}
\hskip 0cm
\includegraphics[width=8cm]{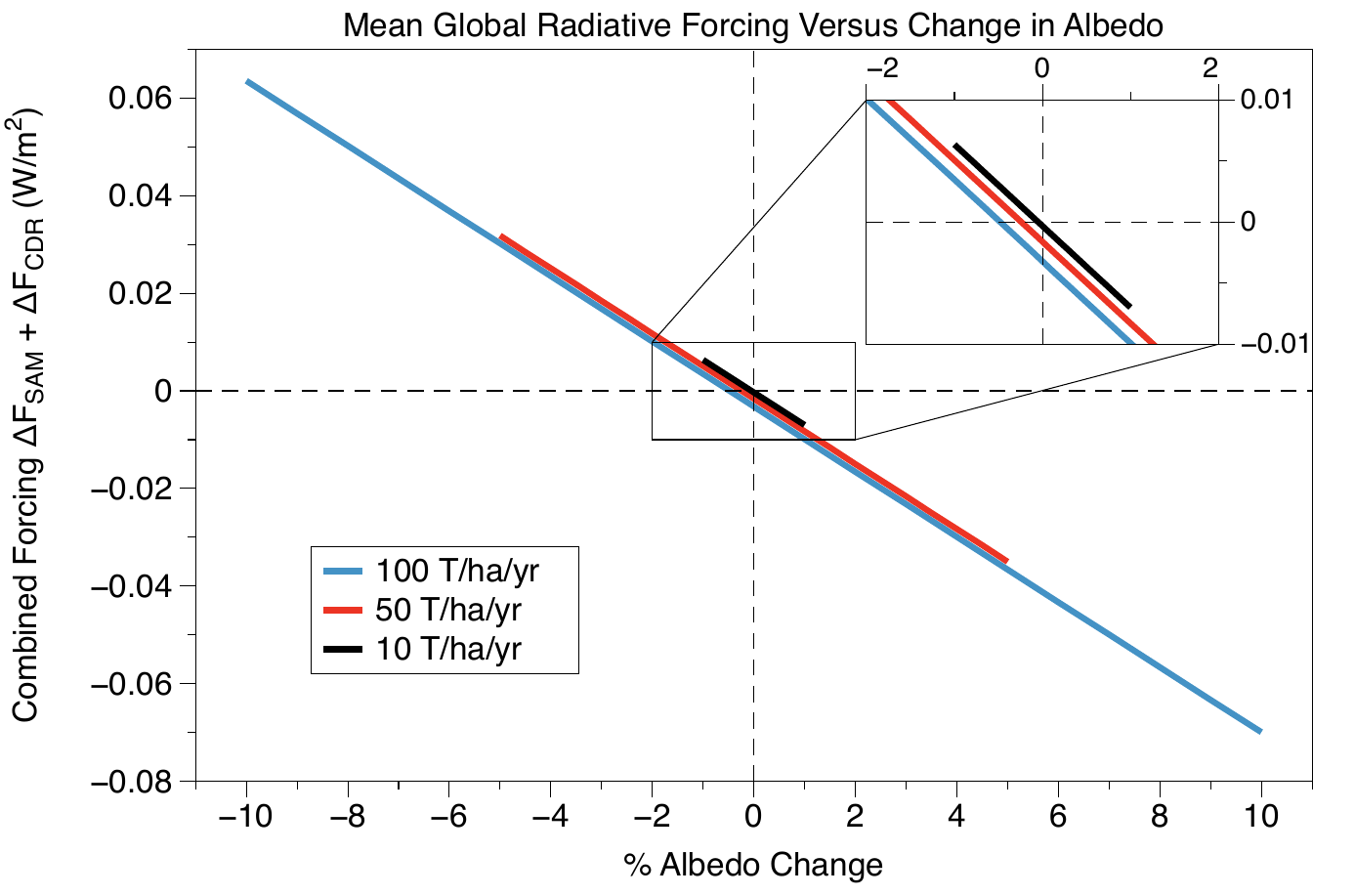}
\end{center}
\caption[x]{(Left) Time evolution of the radiative forcing from treatment of one hectare of land. The enhanced rock weathering (ERW) assumes annual applications of 50 tons of crushed minerals that absorb atmospheric CO\sub{2} at an efficiency of either 5\% or 10\% the mineral mass.  The surface albedo modification (SAM) is from an assumed constant-in-time increase in albedo due to the soil amendments by $0.1$.  Note that ERW only catches up to SAM after multiple decades. (Right) Global mean radiative forcing after 1 year from ERW and SAM on $10^6$ km$^2$ of land versus albedo change for three application rates assuming a 5\% efficiency. As the albedo response to ERW depends on many factors, lines here represent possible values ranging from positive changes (cooling) to negative (warming).}
\label{forcing}
\end{figure}

A spreadsheet model of ERW and SAM \footnote{See the Supplemental Materials for the (Google Sheets) model at URL https://tinyurl.com/ydsdwr5w} permits parameters to be easily adjusted.  The default values correspond to the case discussed above.  More realistically we expect typical changes in the albedo of order 1\% to 10\% leading to SAM radiative forcing that dominates over ERW for years or decades after treatment until sufficient CO\sub{2} has been absorbed as shown in Figure \ref{forcing} (left panel). The right panel of the Figure also shows that SAM is dominant in the short term, and can be either cooling or warming depending on the sign of the albedo change.

\section{Scientific Questions}
\label{Questions}
We are unaware of any prior work done to measure albedo changes due to soil amendments with crushed silicate rock.  
Given the lack of knowledge of possible albedo changes, laboratory and field measurements are a first priority.  Measurements of surface reflectance can be performed with portable albedometers built from pairs of calibrated pyranometers, with one directed to the ground below and another aimed at the sky above, to measure the ratio of the outgoing to incoming visible radiation.  

Multiple albedo measurements will need to be made in each plot to understand statistical variations in albedo, similar to tracking carbon dioxide consumption \cite{Reershemius.2023,Derry2024,Mooshammer2024}, and therefore reduce uncertainty in the measurement of the mean albedo. Once large enough areas of land or sea are employed for CDR, albedo changes could be monitored from air or even from space. The initial focus should be on answering some basic questions:

\begin{itemize}
    
    \item How does the albedo depend on time of day (angle of the sun), time of year, soil moisture, depth of soil tilling, soil biogeochemistry, and crop cover? 

    \item Can albedo be modeled as a weighted average of the separate soil and mineral albedos? Is the weighting factor the fractional mass density, fractional volume, fractional area, or something else? What is the dependence on the mineral grain size?

    \item Do iron and other heavier elements always reduce albedo?  Can mixtures be chosen to simultaneously optimize high ERW rates, high albedo, and mineral availability while at the same time minimizing heavy elements such as chromium and nickel that can contaminate soil?

    \item Does tilling mix the rock dust uniformly in the soil or does a significant portion remain on the top?
    
    \item How does the albedo evolve with repeated soil amendments and tilling? How does it change as the crushed rock weathers?
    
    \item Over what time scale does ERW catch up to SAM in its radiative impact on Earth's climate?
\end{itemize}
Some questions related to mCDR are mentioned in Sec. \ref{mCDR}.

\section{Conclusion}
\label{Conclusion}
Radiative forcing from albedo changes must be considered for any type of CDR that alters large surface land or ocean surface areas. ERW and mCDR are tightly linked to SRM.   
A program of laboratory and field albedo measurements  is needed to quantify SAM. We have identified some of the many open questions that require addressing in the near term before large-scale deployments proceed.  Climate modeling of different spatial and temporal patterns of planned global-scale SAM is also needed to better quantify climate change due to radiative forcing  beyond mean global temperature, such as possible changes in precipitation patterns \cite{Cheng.2023}, localized heating \cite{Cheng.2024} and cloud feedback from SAM. 

\section{Acknowledgments}
JBM thanks Bhavna Arora, Baylor Fox-Kemper and Wati Taylor for helpful discussions. DEI thanks Tyler Kukla and Louis Derry for discussions. We also thank the Brown University Initiative for Sustainable Energy for support.


\begin{thebibliography}{39}%
\makeatletter
\providecommand \@ifxundefined [1]{%
 \@ifx{#1\undefined}
}%
\providecommand \@ifnum [1]{%
 \ifnum #1\expandafter \@firstoftwo
 \else \expandafter \@secondoftwo
 \fi
}%
\providecommand \@ifx [1]{%
 \ifx #1\expandafter \@firstoftwo
 \else \expandafter \@secondoftwo
 \fi
}%
\providecommand \natexlab [1]{#1}%
\providecommand \enquote  [1]{``#1''}%
\providecommand \bibnamefont  [1]{#1}%
\providecommand \bibfnamefont [1]{#1}%
\providecommand \citenamefont [1]{#1}%
\providecommand \href@noop [0]{\@secondoftwo}%
\providecommand \href [0]{\begingroup \@sanitize@url \@href}%
\providecommand \@href[1]{\@@startlink{#1}\@@href}%
\providecommand \@@href[1]{\endgroup#1\@@endlink}%
\providecommand \@sanitize@url [0]{\catcode `\\12\catcode `\$12\catcode
  `\&12\catcode `\#12\catcode `\^12\catcode `\_12\catcode `\%12\relax}%
\providecommand \@@startlink[1]{}%
\providecommand \@@endlink[0]{}%
\providecommand \url  [0]{\begingroup\@sanitize@url \@url }%
\providecommand \@url [1]{\endgroup\@href {#1}{\urlprefix }}%
\providecommand \urlprefix  [0]{URL }%
\providecommand \Eprint [0]{\href }%
\providecommand \doibase [0]{https://doi.org/}%
\providecommand \selectlanguage [0]{\@gobble}%
\providecommand \bibinfo  [0]{\@secondoftwo}%
\providecommand \bibfield  [0]{\@secondoftwo}%
\providecommand \translation [1]{[#1]}%
\providecommand \BibitemOpen [0]{}%
\providecommand \bibitemStop [0]{}%
\providecommand \bibitemNoStop [0]{.\EOS\space}%
\providecommand \EOS [0]{\spacefactor3000\relax}%
\providecommand \BibitemShut  [1]{\csname bibitem#1\endcsname}%
\let\auto@bib@innerbib\@empty
\bibitem [{\citenamefont {{American Geophysical Union}}(2024)}]{AGU.2024}%
  \BibitemOpen
  \bibfield  {author} {\bibinfo {author} {\bibnamefont {{American Geophysical
  Union}}},\ }\href {https://doi.org/10.22541/essoar.172917365.53105072/v1}
  {\bibinfo {title} {Ethical framework principles for climate intervention
  research}},\ \bibinfo {howpublished} {ESS Open Archive} (\bibinfo {year}
  {2024})\BibitemShut {NoStop}%
\bibitem [{\citenamefont {Taylor}\ \emph {et~al.}(2024)\citenamefont {Taylor},
  \citenamefont {Marston}, \citenamefont {Rosner},\ and\ \citenamefont
  {Wurtele}}]{Taylor.2024}%
  \BibitemOpen
  \bibfield  {author} {\bibinfo {author} {\bibfnamefont {W.}~\bibnamefont
  {Taylor}}, \bibinfo {author} {\bibfnamefont {J.~B.}\ \bibnamefont {Marston}},
  \bibinfo {author} {\bibfnamefont {R.}~\bibnamefont {Rosner}},\ and\ \bibinfo
  {author} {\bibfnamefont {J.}~\bibnamefont {Wurtele}},\ }\bibfield  {title}
  {\bibinfo {title} {{Atmospheric Carbon Dioxide Removal (CDR) -- a Physical
  Science Perspective}},\ }\href@noop {} {\bibfield  {journal} {\bibinfo
  {journal} {PRX Energy (to appear)}\ } (\bibinfo {year} {2024})}\BibitemShut
  {NoStop}%
\bibitem [{\citenamefont {Mykleby}\ \emph {et~al.}(2017)\citenamefont
  {Mykleby}, \citenamefont {Snyder},\ and\ \citenamefont
  {Twine}}]{Mykleby.2017}%
  \BibitemOpen
  \bibfield  {author} {\bibinfo {author} {\bibfnamefont {P.~M.}\ \bibnamefont
  {Mykleby}}, \bibinfo {author} {\bibfnamefont {P.~K.}\ \bibnamefont
  {Snyder}},\ and\ \bibinfo {author} {\bibfnamefont {T.~E.}\ \bibnamefont
  {Twine}},\ }\bibfield  {title} {\bibinfo {title} {{Quantifying the
  trade?off between carbon sequestration and albedo in midlatitude and
  high?latitude North American forests}},\ }\href
  {https://doi.org/10.1002/2016gl071459} {\bibfield  {journal} {\bibinfo
  {journal} {Geophysical Research Letters}\ }\textbf {\bibinfo {volume} {44}},\
  \bibinfo {pages} {2493} (\bibinfo {year} {2017})}\BibitemShut {NoStop}%
\bibitem [{\citenamefont {Weber}\ \emph {et~al.}(2024)\citenamefont {Weber},
  \citenamefont {King}, \citenamefont {Abraham}, \citenamefont {Grosvenor},
  \citenamefont {Smith}, \citenamefont {Shin}, \citenamefont {Lawrence},
  \citenamefont {Roe}, \citenamefont {Beerling},\ and\ \citenamefont
  {Martin}}]{Weber.2024}%
  \BibitemOpen
  \bibfield  {author} {\bibinfo {author} {\bibfnamefont {J.}~\bibnamefont
  {Weber}}, \bibinfo {author} {\bibfnamefont {J.~A.}\ \bibnamefont {King}},
  \bibinfo {author} {\bibfnamefont {N.~L.}\ \bibnamefont {Abraham}}, \bibinfo
  {author} {\bibfnamefont {D.~P.}\ \bibnamefont {Grosvenor}}, \bibinfo {author}
  {\bibfnamefont {C.~J.}\ \bibnamefont {Smith}}, \bibinfo {author}
  {\bibfnamefont {Y.~M.}\ \bibnamefont {Shin}}, \bibinfo {author}
  {\bibfnamefont {P.}~\bibnamefont {Lawrence}}, \bibinfo {author}
  {\bibfnamefont {S.}~\bibnamefont {Roe}}, \bibinfo {author} {\bibfnamefont
  {D.~J.}\ \bibnamefont {Beerling}},\ and\ \bibinfo {author} {\bibfnamefont
  {M.~V.}\ \bibnamefont {Martin}},\ }\bibfield  {title} {\bibinfo {title}
  {{Chemistry-albedo feedbacks offset up to a third of forestation's CO2
  removal benefits}},\ }\href {https://doi.org/10.1126/science.adg6196}
  {\bibfield  {journal} {\bibinfo  {journal} {Science}\ }\textbf {\bibinfo
  {volume} {383}},\ \bibinfo {pages} {860} (\bibinfo {year}
  {2024})}\BibitemShut {NoStop}%
\bibitem [{\citenamefont {Kristensen}\ \emph {et~al.}(2024)\citenamefont
  {Kristensen}, \citenamefont {Barbero-Palacios}, \citenamefont {Barrio},
  \citenamefont {Jacobsen}, \citenamefont {Kerby}, \citenamefont
  {L{\'o}pez-Blanco}, \citenamefont {Malhi}, \citenamefont {Moullec},
  \citenamefont {Mueller}, \citenamefont {Post}, \citenamefont {Raundrup},\
  and\ \citenamefont {Macias-Fauria}}]{Kristensen.2024}%
  \BibitemOpen
  \bibfield  {author} {\bibinfo {author} {\bibfnamefont {J.~{\AA}.}\
  \bibnamefont {Kristensen}}, \bibinfo {author} {\bibfnamefont
  {L.}~\bibnamefont {Barbero-Palacios}}, \bibinfo {author} {\bibfnamefont
  {I.~C.}\ \bibnamefont {Barrio}}, \bibinfo {author} {\bibfnamefont {I.~B.~D.}\
  \bibnamefont {Jacobsen}}, \bibinfo {author} {\bibfnamefont {J.~T.}\
  \bibnamefont {Kerby}}, \bibinfo {author} {\bibfnamefont {E.}~\bibnamefont
  {L{\'o}pez-Blanco}}, \bibinfo {author} {\bibfnamefont {Y.}~\bibnamefont
  {Malhi}}, \bibinfo {author} {\bibfnamefont {M.~L.}\ \bibnamefont {Moullec}},
  \bibinfo {author} {\bibfnamefont {C.~W.}\ \bibnamefont {Mueller}}, \bibinfo
  {author} {\bibfnamefont {E.}~\bibnamefont {Post}}, \bibinfo {author}
  {\bibfnamefont {K.}~\bibnamefont {Raundrup}},\ and\ \bibinfo {author}
  {\bibfnamefont {M.}~\bibnamefont {Macias-Fauria}},\ }\bibfield  {title}
  {\bibinfo {title} {{Tree planting is no climate solution at northern high
  latitudes}},\ }\href {https://doi.org/10.1038/s41561-024-01573-4} {\bibfield
  {journal} {\bibinfo  {journal} {Nature Geoscience}\ }\textbf {\bibinfo
  {volume} {17}},\ \bibinfo {pages} {1087} (\bibinfo {year}
  {2024})}\BibitemShut {NoStop}%
\bibitem [{\citenamefont {Beerling}\ \emph {et~al.}(2024)\citenamefont
  {Beerling}, \citenamefont {Epihov}, \citenamefont {Kantola}, \citenamefont
  {Masters}, \citenamefont {Reershemius}, \citenamefont {Planavsky},
  \citenamefont {Reinhard}, \citenamefont {Jordan}, \citenamefont {Thorne},
  \citenamefont {Weber}, \citenamefont {Martin}, \citenamefont {Freckleton},
  \citenamefont {Hartley}, \citenamefont {James}, \citenamefont {Pearce},
  \citenamefont {DeLucia},\ and\ \citenamefont {Banwart}}]{Beerling.2024}%
  \BibitemOpen
  \bibfield  {author} {\bibinfo {author} {\bibfnamefont {D.~J.}\ \bibnamefont
  {Beerling}}, \bibinfo {author} {\bibfnamefont {D.~Z.}\ \bibnamefont
  {Epihov}}, \bibinfo {author} {\bibfnamefont {I.~B.}\ \bibnamefont {Kantola}},
  \bibinfo {author} {\bibfnamefont {M.~D.}\ \bibnamefont {Masters}}, \bibinfo
  {author} {\bibfnamefont {T.}~\bibnamefont {Reershemius}}, \bibinfo {author}
  {\bibfnamefont {N.~J.}\ \bibnamefont {Planavsky}}, \bibinfo {author}
  {\bibfnamefont {C.~T.}\ \bibnamefont {Reinhard}}, \bibinfo {author}
  {\bibfnamefont {J.~S.}\ \bibnamefont {Jordan}}, \bibinfo {author}
  {\bibfnamefont {S.~J.}\ \bibnamefont {Thorne}}, \bibinfo {author}
  {\bibfnamefont {J.}~\bibnamefont {Weber}}, \bibinfo {author} {\bibfnamefont
  {M.~V.}\ \bibnamefont {Martin}}, \bibinfo {author} {\bibfnamefont {R.~P.}\
  \bibnamefont {Freckleton}}, \bibinfo {author} {\bibfnamefont {S.~E.}\
  \bibnamefont {Hartley}}, \bibinfo {author} {\bibfnamefont {R.~H.}\
  \bibnamefont {James}}, \bibinfo {author} {\bibfnamefont {C.~R.}\ \bibnamefont
  {Pearce}}, \bibinfo {author} {\bibfnamefont {E.~H.}\ \bibnamefont
  {DeLucia}},\ and\ \bibinfo {author} {\bibfnamefont {S.~A.}\ \bibnamefont
  {Banwart}},\ }\bibfield  {title} {\bibinfo {title} {{Enhanced weathering in
  the US Corn Belt delivers carbon removal with agronomic benefits}},\ }\href
  {https://doi.org/10.1073/pnas.2319436121} {\bibfield  {journal} {\bibinfo
  {journal} {Proceedings of the National Academy of Sciences}\ }\textbf
  {\bibinfo {volume} {121}},\ \bibinfo {pages} {e2319436121} (\bibinfo {year}
  {2024})}\BibitemShut {NoStop}%
\bibitem [{\citenamefont {Szerszynski}\ \emph {et~al.}(2012)\citenamefont
  {Szerszynski}, \citenamefont {Kearnes}, \citenamefont {Macnaghten},
  \citenamefont {Owen},\ and\ \citenamefont {Stilgoe}}]{Szerszynski.2013}%
  \BibitemOpen
  \bibfield  {author} {\bibinfo {author} {\bibfnamefont {B.}~\bibnamefont
  {Szerszynski}}, \bibinfo {author} {\bibfnamefont {M.}~\bibnamefont
  {Kearnes}}, \bibinfo {author} {\bibfnamefont {P.}~\bibnamefont {Macnaghten}},
  \bibinfo {author} {\bibfnamefont {R.}~\bibnamefont {Owen}},\ and\ \bibinfo
  {author} {\bibfnamefont {J.}~\bibnamefont {Stilgoe}},\ }\bibfield  {title}
  {\bibinfo {title} {{Why Solar Radiation Management Geoengineering and
  Democracy Won't Mix}},\ }\href {https://doi.org/10.1068/a45649} {\bibfield
  {journal} {\bibinfo  {journal} {Environment and Planning A}\ }\textbf
  {\bibinfo {volume} {45}},\ \bibinfo {pages} {2809} (\bibinfo {year}
  {2012})}\BibitemShut {NoStop}%
\bibitem [{\citenamefont {Contzen}\ \emph {et~al.}(2024)\citenamefont
  {Contzen}, \citenamefont {Perlaviciute}, \citenamefont {Steg}, \citenamefont
  {Reckels}, \citenamefont {Alves}, \citenamefont {Bidwell}, \citenamefont
  {B{\"o}hm}, \citenamefont {Bonaiuto}, \citenamefont {Chou}, \citenamefont
  {Corral-Verdugo}, \citenamefont {Dessi}, \citenamefont {Dietz}, \citenamefont
  {Doran}, \citenamefont {Eul{\'a}lio}, \citenamefont {Fielding}, \citenamefont
  {G{\'o}mez-Rom{\'a}n}, \citenamefont {Granskaya}, \citenamefont {Gurikova},
  \citenamefont {Hern{\'a}ndez}, \citenamefont {Kabakova}, \citenamefont {Lee},
  \citenamefont {Li}, \citenamefont {Lima}, \citenamefont {Liu}, \citenamefont
  {Lu{\'\i}s}, \citenamefont {Muinos}, \citenamefont {Ogunbode}, \citenamefont
  {Ortiz}, \citenamefont {Pidgeon}, \citenamefont {Pitt}, \citenamefont
  {Rahimi}, \citenamefont {Revokatova}, \citenamefont {Reyna}, \citenamefont
  {Schuitema}, \citenamefont {Shwom}, \citenamefont {Yalcinkaya}, \citenamefont
  {Spence},\ and\ \citenamefont {S{\"u}tterlin}}]{Contzen.2024}%
  \BibitemOpen
  \bibfield  {author} {\bibinfo {author} {\bibfnamefont {N.}~\bibnamefont
  {Contzen}}, \bibinfo {author} {\bibfnamefont {G.}~\bibnamefont
  {Perlaviciute}}, \bibinfo {author} {\bibfnamefont {L.}~\bibnamefont {Steg}},
  \bibinfo {author} {\bibfnamefont {S.~C.}\ \bibnamefont {Reckels}}, \bibinfo
  {author} {\bibfnamefont {S.}~\bibnamefont {Alves}}, \bibinfo {author}
  {\bibfnamefont {D.}~\bibnamefont {Bidwell}}, \bibinfo {author} {\bibfnamefont
  {G.}~\bibnamefont {B{\"o}hm}}, \bibinfo {author} {\bibfnamefont
  {M.}~\bibnamefont {Bonaiuto}}, \bibinfo {author} {\bibfnamefont {L.-F.}\
  \bibnamefont {Chou}}, \bibinfo {author} {\bibfnamefont {V.}~\bibnamefont
  {Corral-Verdugo}}, \bibinfo {author} {\bibfnamefont {F.}~\bibnamefont
  {Dessi}}, \bibinfo {author} {\bibfnamefont {T.}~\bibnamefont {Dietz}},
  \bibinfo {author} {\bibfnamefont {R.}~\bibnamefont {Doran}}, \bibinfo
  {author} {\bibfnamefont {M.~d.~C.}\ \bibnamefont {Eul{\'a}lio}}, \bibinfo
  {author} {\bibfnamefont {K.}~\bibnamefont {Fielding}}, \bibinfo {author}
  {\bibfnamefont {C.}~\bibnamefont {G{\'o}mez-Rom{\'a}n}}, \bibinfo {author}
  {\bibfnamefont {J.~V.}\ \bibnamefont {Granskaya}}, \bibinfo {author}
  {\bibfnamefont {T.}~\bibnamefont {Gurikova}}, \bibinfo {author}
  {\bibfnamefont {B.}~\bibnamefont {Hern{\'a}ndez}}, \bibinfo {author}
  {\bibfnamefont {M.~P.}\ \bibnamefont {Kabakova}}, \bibinfo {author}
  {\bibfnamefont {C.-Y.}\ \bibnamefont {Lee}}, \bibinfo {author} {\bibfnamefont
  {F.}~\bibnamefont {Li}}, \bibinfo {author} {\bibfnamefont {M.~L.}\
  \bibnamefont {Lima}}, \bibinfo {author} {\bibfnamefont {L.}~\bibnamefont
  {Liu}}, \bibinfo {author} {\bibfnamefont {S.}~\bibnamefont {Lu{\'\i}s}},
  \bibinfo {author} {\bibfnamefont {G.}~\bibnamefont {Muinos}}, \bibinfo
  {author} {\bibfnamefont {C.~A.}\ \bibnamefont {Ogunbode}}, \bibinfo {author}
  {\bibfnamefont {M.~V.}\ \bibnamefont {Ortiz}}, \bibinfo {author}
  {\bibfnamefont {N.}~\bibnamefont {Pidgeon}}, \bibinfo {author} {\bibfnamefont
  {M.~A.}\ \bibnamefont {Pitt}}, \bibinfo {author} {\bibfnamefont
  {L.}~\bibnamefont {Rahimi}}, \bibinfo {author} {\bibfnamefont
  {A.}~\bibnamefont {Revokatova}}, \bibinfo {author} {\bibfnamefont
  {C.}~\bibnamefont {Reyna}}, \bibinfo {author} {\bibfnamefont
  {G.}~\bibnamefont {Schuitema}}, \bibinfo {author} {\bibfnamefont
  {R.}~\bibnamefont {Shwom}}, \bibinfo {author} {\bibfnamefont {N.~S.}\
  \bibnamefont {Yalcinkaya}}, \bibinfo {author} {\bibfnamefont
  {E.}~\bibnamefont {Spence}},\ and\ \bibinfo {author} {\bibfnamefont
  {B.}~\bibnamefont {S{\"u}tterlin}},\ }\bibfield  {title} {\bibinfo {title}
  {Public opinion about solar radiation management: A cross-cultural study in
  20 countries around the world},\ }\href
  {https://doi.org/10.1007/s10584-024-03708-3} {\bibfield  {journal} {\bibinfo
  {journal} {Climatic Change}\ }\textbf {\bibinfo {volume} {177}},\ \bibinfo
  {pages} {65} (\bibinfo {year} {2024})}\BibitemShut {NoStop}%
\bibitem [{\citenamefont {Sugiyama}\ \emph {et~al.}(2024)\citenamefont
  {Sugiyama}, \citenamefont {Asayama}, \citenamefont {Kosugi}, \citenamefont
  {Ishii},\ and\ \citenamefont {Watanabe}}]{Sugiyama.2024}%
  \BibitemOpen
  \bibfield  {author} {\bibinfo {author} {\bibfnamefont {M.}~\bibnamefont
  {Sugiyama}}, \bibinfo {author} {\bibfnamefont {S.}~\bibnamefont {Asayama}},
  \bibinfo {author} {\bibfnamefont {T.}~\bibnamefont {Kosugi}}, \bibinfo
  {author} {\bibfnamefont {A.}~\bibnamefont {Ishii}},\ and\ \bibinfo {author}
  {\bibfnamefont {S.}~\bibnamefont {Watanabe}},\ }\bibfield  {title} {\bibinfo
  {title} {Public attitude toward solar radiation modification: results of a
  two-scenario online survey on perception in four asia--pacific countries},\
  }\bibfield  {journal} {\bibinfo  {journal} {Sustainability Science}\ }\href
  {https://doi.org/10.1007/s11625-024-01520-7} {10.1007/s11625-024-01520-7}
  (\bibinfo {year} {2024})\BibitemShut {NoStop}%
\bibitem [{\citenamefont {Flavelle}(2024)}]{Flavelle.2024}%
  \BibitemOpen
  \bibfield  {author} {\bibinfo {author} {\bibfnamefont {C.}~\bibnamefont
  {Flavelle}},\ }\bibfield  {title} {\bibinfo {title} {{The U.S. Is Building an
  Early Warning System to Detect Geoengineering}},\ }\href
  {https://www.nytimes.com/2024/11/28/climate/geoengineering-early-warning-system.html?unlocked_article_code=1.dU4.cDjS.rd1pAF5iXLvW&smid=url-share}
  {\bibfield  {journal} {\bibinfo  {journal} {New York Times}\ } (\bibinfo
  {year} {2024})}\BibitemShut {NoStop}%
\bibitem [{\citenamefont {Rinder}\ and\ \citenamefont {von
  Hagke}(2021)}]{Rinder2021}%
  \BibitemOpen
  \bibfield  {author} {\bibinfo {author} {\bibfnamefont {T.}~\bibnamefont
  {Rinder}}\ and\ \bibinfo {author} {\bibfnamefont {C.}~\bibnamefont {von
  Hagke}},\ }\bibfield  {title} {\bibinfo {title} {The influence of particle
  size on the potential of enhanced basalt weathering for carbon dioxide
  removal - insights from a regional assessment},\ }\href
  {https://doi.org/10.1016/j.jclepro.2021.128178} {\bibfield  {journal}
  {\bibinfo  {journal} {Journal of Cleaner Production}\ }\textbf {\bibinfo
  {volume} {315}},\ \bibinfo {pages} {128178} (\bibinfo {year}
  {2021})}\BibitemShut {NoStop}%
\bibitem [{\citenamefont {Beerling}\ \emph {et~al.}(2020)\citenamefont
  {Beerling}, \citenamefont {Kantzas}, \citenamefont {Lomas}, \citenamefont
  {Wade}, \citenamefont {Eufrasio}, \citenamefont {Renforth}, \citenamefont
  {Sarkar}, \citenamefont {Andrews}, \citenamefont {James}, \citenamefont
  {Pearce}, \citenamefont {Mercure}, \citenamefont {Pollitt}, \citenamefont
  {Holden}, \citenamefont {Edwards}, \citenamefont {Khanna}, \citenamefont
  {Koh}, \citenamefont {Quegan}, \citenamefont {Pidgeon}, \citenamefont
  {Janssens}, \citenamefont {Hansen},\ and\ \citenamefont
  {Banwart}}]{Beerling.2020}%
  \BibitemOpen
  \bibfield  {author} {\bibinfo {author} {\bibfnamefont {D.~J.}\ \bibnamefont
  {Beerling}}, \bibinfo {author} {\bibfnamefont {E.~P.}\ \bibnamefont
  {Kantzas}}, \bibinfo {author} {\bibfnamefont {M.~R.}\ \bibnamefont {Lomas}},
  \bibinfo {author} {\bibfnamefont {P.}~\bibnamefont {Wade}}, \bibinfo {author}
  {\bibfnamefont {R.~M.}\ \bibnamefont {Eufrasio}}, \bibinfo {author}
  {\bibfnamefont {P.}~\bibnamefont {Renforth}}, \bibinfo {author}
  {\bibfnamefont {B.}~\bibnamefont {Sarkar}}, \bibinfo {author} {\bibfnamefont
  {M.~G.}\ \bibnamefont {Andrews}}, \bibinfo {author} {\bibfnamefont {R.~H.}\
  \bibnamefont {James}}, \bibinfo {author} {\bibfnamefont {C.~R.}\ \bibnamefont
  {Pearce}}, \bibinfo {author} {\bibfnamefont {J.-F.}\ \bibnamefont {Mercure}},
  \bibinfo {author} {\bibfnamefont {H.}~\bibnamefont {Pollitt}}, \bibinfo
  {author} {\bibfnamefont {P.~B.}\ \bibnamefont {Holden}}, \bibinfo {author}
  {\bibfnamefont {N.~R.}\ \bibnamefont {Edwards}}, \bibinfo {author}
  {\bibfnamefont {M.}~\bibnamefont {Khanna}}, \bibinfo {author} {\bibfnamefont
  {L.}~\bibnamefont {Koh}}, \bibinfo {author} {\bibfnamefont {S.}~\bibnamefont
  {Quegan}}, \bibinfo {author} {\bibfnamefont {N.~F.}\ \bibnamefont {Pidgeon}},
  \bibinfo {author} {\bibfnamefont {I.~A.}\ \bibnamefont {Janssens}}, \bibinfo
  {author} {\bibfnamefont {J.}~\bibnamefont {Hansen}},\ and\ \bibinfo {author}
  {\bibfnamefont {S.~A.}\ \bibnamefont {Banwart}},\ }\bibfield  {title}
  {\bibinfo {title} {{Potential for large-scale CO2 removal via enhanced rock
  weathering with croplands}},\ }\href@noop {} {\bibfield  {journal} {\bibinfo
  {journal} {Nature}\ }\textbf {\bibinfo {volume} {583}},\ \bibinfo {pages}
  {242} (\bibinfo {year} {2020})}\BibitemShut {NoStop}%
\bibitem [{\citenamefont {Oh}\ and\ \citenamefont {Raymond}(2006)}]{Oh2006}%
  \BibitemOpen
  \bibfield  {author} {\bibinfo {author} {\bibfnamefont {N.}~\bibnamefont
  {Oh}}\ and\ \bibinfo {author} {\bibfnamefont {P.~A.}\ \bibnamefont
  {Raymond}},\ }\bibfield  {title} {\bibinfo {title} {Contribution of
  agricultural liming to riverine bicarbonate export and co2 sequestration in
  the ohio river basin},\ }\bibfield  {journal} {\bibinfo  {journal} {Global
  Biogeochemical Cycles}\ }\textbf {\bibinfo {volume} {20}},\ \href
  {https://doi.org/10.1029/2005gb002565} {10.1029/2005gb002565} (\bibinfo
  {year} {2006})\BibitemShut {NoStop}%
\bibitem [{\citenamefont {Zhang}\ \emph {et~al.}(2024)\citenamefont {Zhang},
  \citenamefont {Reinhard}, \citenamefont {Liu}, \citenamefont {Kanzaki},\ and\
  \citenamefont {Planavsky}}]{Zhang2024}%
  \BibitemOpen
  \bibfield  {author} {\bibinfo {author} {\bibfnamefont {S.}~\bibnamefont
  {Zhang}}, \bibinfo {author} {\bibfnamefont {C.~T.}\ \bibnamefont {Reinhard}},
  \bibinfo {author} {\bibfnamefont {S.}~\bibnamefont {Liu}}, \bibinfo {author}
  {\bibfnamefont {Y.}~\bibnamefont {Kanzaki}},\ and\ \bibinfo {author}
  {\bibfnamefont {N.~J.}\ \bibnamefont {Planavsky}},\ }\bibfield  {title}
  {\bibinfo {title} {A framework for modeling carbon loss from rivers following
  terrestrial enhanced weathering},\ }\bibfield  {journal} {\bibinfo  {journal}
  {Environmental Research Letters}\ }\href
  {https://doi.org/10.1088/1748-9326/ada398} {10.1088/1748-9326/ada398}
  (\bibinfo {year} {2024})\BibitemShut {NoStop}%
\bibitem [{\citenamefont {Manning}\ \emph {et~al.}(2024)\citenamefont
  {Manning}, \citenamefont {de~Azevedo}, \citenamefont {Zani},\ and\
  \citenamefont {Barneze}}]{Manning2024}%
  \BibitemOpen
  \bibfield  {author} {\bibinfo {author} {\bibfnamefont {D.~A.~C.}\
  \bibnamefont {Manning}}, \bibinfo {author} {\bibfnamefont {A.~C.}\
  \bibnamefont {de~Azevedo}}, \bibinfo {author} {\bibfnamefont {C.~F.}\
  \bibnamefont {Zani}},\ and\ \bibinfo {author} {\bibfnamefont {A.~S.}\
  \bibnamefont {Barneze}},\ }\bibfield  {title} {\bibinfo {title} {Soil carbon
  management and enhanced rock weathering: The separate fates of organic and
  inorganic carbon},\ }\bibfield  {journal} {\bibinfo  {journal} {European
  Journal of Soil Science}\ }\textbf {\bibinfo {volume} {75}},\ \href
  {https://doi.org/10.1111/ejss.13534} {10.1111/ejss.13534} (\bibinfo {year}
  {2024})\BibitemShut {NoStop}%
\bibitem [{\citenamefont {Campbell}\ \emph {et~al.}(2022)\citenamefont
  {Campbell}, \citenamefont {Foteinis}, \citenamefont {Furey}, \citenamefont
  {Hawrot}, \citenamefont {Pike}, \citenamefont {Aeschlimann}, \citenamefont
  {Maesano}, \citenamefont {Reginato}, \citenamefont {Goodwin}, \citenamefont
  {Looger}, \citenamefont {Boyden},\ and\ \citenamefont
  {Renforth}}]{Campbell.2022}%
  \BibitemOpen
  \bibfield  {author} {\bibinfo {author} {\bibfnamefont {J.~S.}\ \bibnamefont
  {Campbell}}, \bibinfo {author} {\bibfnamefont {S.}~\bibnamefont {Foteinis}},
  \bibinfo {author} {\bibfnamefont {V.}~\bibnamefont {Furey}}, \bibinfo
  {author} {\bibfnamefont {O.}~\bibnamefont {Hawrot}}, \bibinfo {author}
  {\bibfnamefont {D.}~\bibnamefont {Pike}}, \bibinfo {author} {\bibfnamefont
  {S.}~\bibnamefont {Aeschlimann}}, \bibinfo {author} {\bibfnamefont {C.~N.}\
  \bibnamefont {Maesano}}, \bibinfo {author} {\bibfnamefont {P.~L.}\
  \bibnamefont {Reginato}}, \bibinfo {author} {\bibfnamefont {D.~R.}\
  \bibnamefont {Goodwin}}, \bibinfo {author} {\bibfnamefont {L.~L.}\
  \bibnamefont {Looger}}, \bibinfo {author} {\bibfnamefont {E.~S.}\
  \bibnamefont {Boyden}},\ and\ \bibinfo {author} {\bibfnamefont
  {P.}~\bibnamefont {Renforth}},\ }\bibfield  {title} {\bibinfo {title}
  {{Geochemical Negative Emissions Technologies: Part I. Review}},\ }\href@noop
  {} {\bibfield  {journal} {\bibinfo  {journal} {Frontiers in Climate}\
  }\textbf {\bibinfo {volume} {4}},\ \bibinfo {pages} {879133} (\bibinfo {year}
  {2022})}\BibitemShut {NoStop}%
\bibitem [{\citenamefont {Breunig}\ \emph {et~al.}(2024)\citenamefont
  {Breunig}, \citenamefont {Fox}, \citenamefont {Domen}, \citenamefont {Kumar},
  \citenamefont {Alves}, \citenamefont {Zhalnina}, \citenamefont
  {Voigtl{\"a}nder}, \citenamefont {Deng}, \citenamefont {Arora},\ and\
  \citenamefont {Nico}}]{Breunig.2024}%
  \BibitemOpen
  \bibfield  {author} {\bibinfo {author} {\bibfnamefont {H.~M.}\ \bibnamefont
  {Breunig}}, \bibinfo {author} {\bibfnamefont {P.}~\bibnamefont {Fox}},
  \bibinfo {author} {\bibfnamefont {J.}~\bibnamefont {Domen}}, \bibinfo
  {author} {\bibfnamefont {R.}~\bibnamefont {Kumar}}, \bibinfo {author}
  {\bibfnamefont {R.~J.~E.}\ \bibnamefont {Alves}}, \bibinfo {author}
  {\bibfnamefont {K.}~\bibnamefont {Zhalnina}}, \bibinfo {author}
  {\bibfnamefont {A.}~\bibnamefont {Voigtl{\"a}nder}}, \bibinfo {author}
  {\bibfnamefont {H.}~\bibnamefont {Deng}}, \bibinfo {author} {\bibfnamefont
  {B.}~\bibnamefont {Arora}},\ and\ \bibinfo {author} {\bibfnamefont
  {P.}~\bibnamefont {Nico}},\ }\bibfield  {title} {\bibinfo {title} {{Life
  cycle impact and cost analysis of quarry materials for land-based enhanced
  weathering in Northern California}},\ }\href
  {https://doi.org/10.1016/j.jclepro.2024.143757} {\bibfield  {journal}
  {\bibinfo  {journal} {Journal of Cleaner Production}\ ,\ \bibinfo {pages}
  {143757}} (\bibinfo {year} {2024})}\BibitemShut {NoStop}%
\bibitem [{\citenamefont {Goll}\ \emph {et~al.}(2021)\citenamefont {Goll},
  \citenamefont {Ciais}, \citenamefont {Amann}, \citenamefont {Buermann},
  \citenamefont {Chang}, \citenamefont {Eker}, \citenamefont {Hartmann},
  \citenamefont {Janssens}, \citenamefont {Li}, \citenamefont {Obersteiner},
  \citenamefont {Penuelas}, \citenamefont {Tanaka},\ and\ \citenamefont
  {Vicca}}]{Goll.2021}%
  \BibitemOpen
  \bibfield  {author} {\bibinfo {author} {\bibfnamefont {D.~S.}\ \bibnamefont
  {Goll}}, \bibinfo {author} {\bibfnamefont {P.}~\bibnamefont {Ciais}},
  \bibinfo {author} {\bibfnamefont {T.}~\bibnamefont {Amann}}, \bibinfo
  {author} {\bibfnamefont {W.}~\bibnamefont {Buermann}}, \bibinfo {author}
  {\bibfnamefont {J.}~\bibnamefont {Chang}}, \bibinfo {author} {\bibfnamefont
  {S.}~\bibnamefont {Eker}}, \bibinfo {author} {\bibfnamefont {J.}~\bibnamefont
  {Hartmann}}, \bibinfo {author} {\bibfnamefont {I.}~\bibnamefont {Janssens}},
  \bibinfo {author} {\bibfnamefont {W.}~\bibnamefont {Li}}, \bibinfo {author}
  {\bibfnamefont {M.}~\bibnamefont {Obersteiner}}, \bibinfo {author}
  {\bibfnamefont {J.}~\bibnamefont {Penuelas}}, \bibinfo {author}
  {\bibfnamefont {K.}~\bibnamefont {Tanaka}},\ and\ \bibinfo {author}
  {\bibfnamefont {S.}~\bibnamefont {Vicca}},\ }\bibfield  {title} {\bibinfo
  {title} {{Potential CO2 removal from enhanced weathering by ecosystem
  responses to powdered rock}},\ }\href@noop {} {\bibfield  {journal} {\bibinfo
   {journal} {Nature Geoscience}\ }\textbf {\bibinfo {volume} {14}},\ \bibinfo
  {pages} {545} (\bibinfo {year} {2021})}\BibitemShut {NoStop}%
\bibitem [{\citenamefont {Welbel}(2023)}]{Welbel.2023}%
  \BibitemOpen
  \bibfield  {author} {\bibinfo {author} {\bibfnamefont {G.}~\bibnamefont
  {Welbel}},\ }\href
  {https://earth.yale.edu/sites/default/files/files/SeniorEssays/Welbel_Thesis_Final.pdf}
  {\bibinfo {title} {{Enhanced Rock Weathering in Agricultural Settings: Field
  Trial at Zumwalt Acres in Sheldon, IL}}} (\bibinfo {year} {2023})\BibitemShut
  {NoStop}%
\bibitem [{\citenamefont {Haque}\ \emph {et~al.}(2020)\citenamefont {Haque},
  \citenamefont {Santos},\ and\ \citenamefont {Chiang}}]{haque2020optimizing}%
  \BibitemOpen
  \bibfield  {author} {\bibinfo {author} {\bibfnamefont {F.}~\bibnamefont
  {Haque}}, \bibinfo {author} {\bibfnamefont {R.~M.}\ \bibnamefont {Santos}},\
  and\ \bibinfo {author} {\bibfnamefont {Y.~W.}\ \bibnamefont {Chiang}},\
  }\bibfield  {title} {\bibinfo {title} {Optimizing inorganic carbon
  sequestration and crop yield with wollastonite soil amendment in a microplot
  study},\ }\href@noop {} {\bibfield  {journal} {\bibinfo  {journal} {Frontiers
  in plant science}\ }\textbf {\bibinfo {volume} {11}},\ \bibinfo {pages}
  {1012} (\bibinfo {year} {2020})}\BibitemShut {NoStop}%
\bibitem [{\citenamefont {Wood}\ \emph {et~al.}(2023)\citenamefont {Wood},
  \citenamefont {Harrison},\ and\ \citenamefont {Power}}]{wood2023impacts}%
  \BibitemOpen
  \bibfield  {author} {\bibinfo {author} {\bibfnamefont {C.}~\bibnamefont
  {Wood}}, \bibinfo {author} {\bibfnamefont {A.~L.}\ \bibnamefont {Harrison}},\
  and\ \bibinfo {author} {\bibfnamefont {I.~M.}\ \bibnamefont {Power}},\
  }\bibfield  {title} {\bibinfo {title} {Impacts of dissolved phosphorus and
  soil-mineral-fluid interactions on co2 removal through enhanced weathering of
  wollastonite in soils},\ }\href@noop {} {\bibfield  {journal} {\bibinfo
  {journal} {Applied Geochemistry}\ }\textbf {\bibinfo {volume} {148}},\
  \bibinfo {pages} {105511} (\bibinfo {year} {2023})}\BibitemShut {NoStop}%
\bibitem [{\citenamefont {Bai}\ \emph {et~al.}(2019)\citenamefont {Bai},
  \citenamefont {Huang}, \citenamefont {Ren}, \citenamefont {Coyne},
  \citenamefont {Jacinthe}, \citenamefont {Tao}, \citenamefont {Hui},
  \citenamefont {Yang},\ and\ \citenamefont {Matocha}}]{Bai.2019}%
  \BibitemOpen
  \bibfield  {author} {\bibinfo {author} {\bibfnamefont {X.}~\bibnamefont
  {Bai}}, \bibinfo {author} {\bibfnamefont {Y.}~\bibnamefont {Huang}}, \bibinfo
  {author} {\bibfnamefont {W.}~\bibnamefont {Ren}}, \bibinfo {author}
  {\bibfnamefont {M.}~\bibnamefont {Coyne}}, \bibinfo {author} {\bibfnamefont
  {P.-A.}\ \bibnamefont {Jacinthe}}, \bibinfo {author} {\bibfnamefont
  {B.}~\bibnamefont {Tao}}, \bibinfo {author} {\bibfnamefont {D.}~\bibnamefont
  {Hui}}, \bibinfo {author} {\bibfnamefont {J.}~\bibnamefont {Yang}},\ and\
  \bibinfo {author} {\bibfnamefont {C.}~\bibnamefont {Matocha}},\ }\bibfield
  {title} {\bibinfo {title} {Responses of soil carbon sequestration to
  climate-smart agriculture practices: A meta-analysis},\ }\href
  {https://doi.org/10.1111/gcb.14658} {\bibfield  {journal} {\bibinfo
  {journal} {Global Change Biology}\ }\textbf {\bibinfo {volume} {25}},\
  \bibinfo {pages} {2591} (\bibinfo {year} {2019})}\BibitemShut {NoStop}%
\bibitem [{NAS(2022)}]{NASEM.2022}%
  \BibitemOpen
  \href {https://doi.org/10.17226/26278} {\emph {\bibinfo {title} {{A Research
  Strategy for Ocean-based Carbon Dioxide Removal and Sequestration}}}}\
  (\bibinfo  {publisher} {The National Academies Press},\ \bibinfo {address}
  {Washington, DC},\ \bibinfo {year} {2022})\BibitemShut {NoStop}%
\bibitem [{\citenamefont {Lebling}\ \emph {et~al.}(2022)\citenamefont
  {Lebling}, \citenamefont {Northrop}, \citenamefont {McCormick},\ and\
  \citenamefont {Bridgwater}}]{Lebling.2022}%
  \BibitemOpen
  \bibfield  {author} {\bibinfo {author} {\bibfnamefont {K.}~\bibnamefont
  {Lebling}}, \bibinfo {author} {\bibfnamefont {E.}~\bibnamefont {Northrop}},
  \bibinfo {author} {\bibfnamefont {C.}~\bibnamefont {McCormick}},\ and\
  \bibinfo {author} {\bibfnamefont {E.}~\bibnamefont {Bridgwater}},\ }\bibfield
   {title} {\bibinfo {title} {{Towards Responsible and Informed Ocean-Based
  Carbon Dioxide Removal: Research and Governance Priorities}},\ }\href@noop {}
  {\bibfield  {journal} {\bibinfo  {journal} {World Resources Institute}\ }
  (\bibinfo {year} {2022})}\BibitemShut {NoStop}%
\bibitem [{\citenamefont {Solutions}(2023)}]{Buessler.2023}%
  \BibitemOpen
  \bibfield  {author} {\bibinfo {author} {\bibfnamefont {E.~O.~I.}\
  \bibnamefont {Solutions}},\ }\href {https://doi.org/10.1575/1912/67120}
  {\bibinfo {title} {{Paths Forward for Exploring Ocean Iron Fertilization}}}
  (\bibinfo {year} {October 27, 2023})\BibitemShut {NoStop}%
\bibitem [{\citenamefont {Seneviratne}\ \emph {et~al.}(2018)\citenamefont
  {Seneviratne}, \citenamefont {Phipps}, \citenamefont {Pitman}, \citenamefont
  {Hirsch}, \citenamefont {Davin}, \citenamefont {Donat}, \citenamefont
  {Hirschi}, \citenamefont {Lenton}, \citenamefont {Wilhelm},\ and\
  \citenamefont {Kravitz}}]{Seneviratne.2018}%
  \BibitemOpen
  \bibfield  {author} {\bibinfo {author} {\bibfnamefont {S.~I.}\ \bibnamefont
  {Seneviratne}}, \bibinfo {author} {\bibfnamefont {S.~J.}\ \bibnamefont
  {Phipps}}, \bibinfo {author} {\bibfnamefont {A.~J.}\ \bibnamefont {Pitman}},
  \bibinfo {author} {\bibfnamefont {A.~L.}\ \bibnamefont {Hirsch}}, \bibinfo
  {author} {\bibfnamefont {E.~L.}\ \bibnamefont {Davin}}, \bibinfo {author}
  {\bibfnamefont {M.~G.}\ \bibnamefont {Donat}}, \bibinfo {author}
  {\bibfnamefont {M.}~\bibnamefont {Hirschi}}, \bibinfo {author} {\bibfnamefont
  {A.}~\bibnamefont {Lenton}}, \bibinfo {author} {\bibfnamefont
  {M.}~\bibnamefont {Wilhelm}},\ and\ \bibinfo {author} {\bibfnamefont
  {B.}~\bibnamefont {Kravitz}},\ }\bibfield  {title} {\bibinfo {title} {{Land
  radiative management as contributor to regional-scale climate adaptation and
  mitigation}},\ }\href {https://doi.org/10.1038/s41561-017-0057-5} {\bibfield
  {journal} {\bibinfo  {journal} {Nature Geoscience}\ }\textbf {\bibinfo
  {volume} {11}},\ \bibinfo {pages} {88} (\bibinfo {year} {2018})}\BibitemShut
  {NoStop}%
\bibitem [{\citenamefont {{National Research Council}}(2015)}]{NAP.2015}%
  \BibitemOpen
  \bibfield  {author} {\bibinfo {author} {\bibnamefont {{National Research
  Council}}},\ }\href {https://doi.org/10.17226/18988} {\emph {\bibinfo {title}
  {Climate Intervention: Reflecting Sunlight to Cool Earth}}}\ (\bibinfo
  {publisher} {The National Academies Press},\ \bibinfo {address} {Washington,
  DC},\ \bibinfo {year} {2015})\BibitemShut {NoStop}%
\bibitem [{\citenamefont {{Geoengineering Monitor}}(2021)}]{Monitor.2021}%
  \BibitemOpen
  \bibfield  {author} {\bibinfo {author} {\bibnamefont {{Geoengineering
  Monitor}}},\ }\href
  {https://www.geoengineeringmonitor.org/wp-content/uploads/2021/04/surface-albedo-modification.pdf}
  {\bibinfo {title} {{Surface Albedo Modification}}} (\bibinfo {year}
  {2021})\BibitemShut {NoStop}%
\bibitem [{\citenamefont {Johnson}\ \emph {et~al.}(2022)\citenamefont
  {Johnson}, \citenamefont {Manzara}, \citenamefont {Field}, \citenamefont
  {Chamberlin},\ and\ \citenamefont {Sholtz}}]{Johnson.2022}%
  \BibitemOpen
  \bibfield  {author} {\bibinfo {author} {\bibfnamefont {D.}~\bibnamefont
  {Johnson}}, \bibinfo {author} {\bibfnamefont {A.}~\bibnamefont {Manzara}},
  \bibinfo {author} {\bibfnamefont {L.~A.}\ \bibnamefont {Field}}, \bibinfo
  {author} {\bibfnamefont {D.~R.}\ \bibnamefont {Chamberlin}},\ and\ \bibinfo
  {author} {\bibfnamefont {A.}~\bibnamefont {Sholtz}},\ }\bibfield  {title}
  {\bibinfo {title} {{A Controlled Experiment of Surface Albedo Modification to
  Reduce Ice Melt}},\ }\bibfield  {journal} {\bibinfo  {journal} {Earth's
  Future}\ }\textbf {\bibinfo {volume} {10}},\ \href
  {https://doi.org/10.1029/2022ef002883} {10.1029/2022ef002883} (\bibinfo
  {year} {2022})\BibitemShut {NoStop}%
\bibitem [{\citenamefont {Ouyang}\ \emph {et~al.}(2022)\citenamefont {Ouyang},
  \citenamefont {Sciusco}, \citenamefont {Jiao}, \citenamefont {Feron},
  \citenamefont {Lei}, \citenamefont {Li}, \citenamefont {John}, \citenamefont
  {Fan}, \citenamefont {Li}, \citenamefont {Williams}, \citenamefont {Chen},
  \citenamefont {Wang},\ and\ \citenamefont {Chen}}]{Ouyang.2022}%
  \BibitemOpen
  \bibfield  {author} {\bibinfo {author} {\bibfnamefont {Z.}~\bibnamefont
  {Ouyang}}, \bibinfo {author} {\bibfnamefont {P.}~\bibnamefont {Sciusco}},
  \bibinfo {author} {\bibfnamefont {T.}~\bibnamefont {Jiao}}, \bibinfo {author}
  {\bibfnamefont {S.}~\bibnamefont {Feron}}, \bibinfo {author} {\bibfnamefont
  {C.}~\bibnamefont {Lei}}, \bibinfo {author} {\bibfnamefont {F.}~\bibnamefont
  {Li}}, \bibinfo {author} {\bibfnamefont {R.}~\bibnamefont {John}}, \bibinfo
  {author} {\bibfnamefont {P.}~\bibnamefont {Fan}}, \bibinfo {author}
  {\bibfnamefont {X.}~\bibnamefont {Li}}, \bibinfo {author} {\bibfnamefont
  {C.~A.}\ \bibnamefont {Williams}}, \bibinfo {author} {\bibfnamefont
  {G.}~\bibnamefont {Chen}}, \bibinfo {author} {\bibfnamefont {C.}~\bibnamefont
  {Wang}},\ and\ \bibinfo {author} {\bibfnamefont {J.}~\bibnamefont {Chen}},\
  }\bibfield  {title} {\bibinfo {title} {{Albedo changes caused by future
  urbanization contribute to global warming}},\ }\href
  {https://doi.org/10.1038/s41467-022-31558-z} {\bibfield  {journal} {\bibinfo
  {journal} {Nature Communications}\ }\textbf {\bibinfo {volume} {13}},\
  \bibinfo {pages} {3800} (\bibinfo {year} {2022})}\BibitemShut {NoStop}%
\bibitem [{\citenamefont {Cheng}\ \emph {et~al.}(2023)\citenamefont {Cheng},
  \citenamefont {Hu},\ and\ \citenamefont {McColl}}]{Cheng.2023}%
  \BibitemOpen
  \bibfield  {author} {\bibinfo {author} {\bibfnamefont {Y.}~\bibnamefont
  {Cheng}}, \bibinfo {author} {\bibfnamefont {Z.}~\bibnamefont {Hu}},\ and\
  \bibinfo {author} {\bibfnamefont {K.~A.}\ \bibnamefont {McColl}},\ }\bibfield
   {title} {\bibinfo {title} {{Anomalously Darker Land Surfaces Become Wetter
  Due To Mesoscale Circulations}},\ }\bibfield  {journal} {\bibinfo  {journal}
  {Geophysical Research Letters}\ }\textbf {\bibinfo {volume} {50}},\ \href
  {https://doi.org/10.1029/2023gl104137} {10.1029/2023gl104137} (\bibinfo
  {year} {2023})\BibitemShut {NoStop}%
\bibitem [{\citenamefont {Cheng}\ and\ \citenamefont
  {McColl}(2024)}]{Cheng.2024}%
  \BibitemOpen
  \bibfield  {author} {\bibinfo {author} {\bibfnamefont {Y.}~\bibnamefont
  {Cheng}}\ and\ \bibinfo {author} {\bibfnamefont {K.~A.}\ \bibnamefont
  {McColl}},\ }\bibfield  {title} {\bibinfo {title} {{Unexpected Warming From
  Land Radiative Management}},\ }\bibfield  {journal} {\bibinfo  {journal}
  {Geophysical Research Letters}\ }\textbf {\bibinfo {volume} {51}},\ \href
  {https://doi.org/10.1029/2024gl112433} {10.1029/2024gl112433} (\bibinfo
  {year} {2024})\BibitemShut {NoStop}%
\bibitem [{\citenamefont {et~al.}(2021)}]{Masson.2021}%
  \BibitemOpen
  \bibfield  {author} {\bibinfo {author} {\bibfnamefont {V.~M.-D.}\
  \bibnamefont {et~al.}},\ }\href {https://doi.org/10.1017/9781009157896}
  {\emph {\bibinfo {title} {{IPCC, 2021: Climate Change 2021: The Physical
  Science Basis. Contribution of Working Group I to the Sixth Assessment Report
  of the Intergovernmental Panel on Climate Change}}}},\ edited by\ \bibinfo
  {editor} {\bibfnamefont {V.}~\bibnamefont {Masson-Delmotte}}\ and\ \bibinfo
  {editor} {\bibnamefont {et~al.}}\ (\bibinfo  {publisher} {Cambridge
  University Press},\ \bibinfo {address} {Cambridge, UK and New York, NY,
  USA},\ \bibinfo {year} {2021})\BibitemShut {NoStop}%
\bibitem [{\citenamefont {Myhre}\ \emph {et~al.}(1998)\citenamefont {Myhre},
  \citenamefont {Highwood}, \citenamefont {Shine},\ and\ \citenamefont
  {Stordal}}]{Myhre.1998}%
  \BibitemOpen
  \bibfield  {author} {\bibinfo {author} {\bibfnamefont {G.}~\bibnamefont
  {Myhre}}, \bibinfo {author} {\bibfnamefont {E.~J.}\ \bibnamefont {Highwood}},
  \bibinfo {author} {\bibfnamefont {K.~P.}\ \bibnamefont {Shine}},\ and\
  \bibinfo {author} {\bibfnamefont {F.}~\bibnamefont {Stordal}},\ }\bibfield
  {title} {\bibinfo {title} {New estimates of radiative forcing due to well
  mixed greenhouse gases},\ }\href
  {https://doi.org/https://doi.org/10.1029/98GL01908} {\bibfield  {journal}
  {\bibinfo  {journal} {Geophysical Research Letters}\ }\textbf {\bibinfo
  {volume} {25}},\ \bibinfo {pages} {2715} (\bibinfo {year} {1998})},\ \Eprint
  {https://arxiv.org/abs/https://agupubs.onlinelibrary.wiley.com/doi/pdf/10.1029/98GL01908}
  {https://agupubs.onlinelibrary.wiley.com/doi/pdf/10.1029/98GL01908}
  \BibitemShut {NoStop}%
\bibitem [{\citenamefont {Mlynczak}\ \emph {et~al.}(2016)\citenamefont
  {Mlynczak}, \citenamefont {Daniels}, \citenamefont {Kratz}, \citenamefont
  {Feldman}, \citenamefont {Collins}, \citenamefont {Mlawer}, \citenamefont
  {Alvarado}, \citenamefont {Lawler}, \citenamefont {Anderson}, \citenamefont
  {Fahey}, \citenamefont {Hunt},\ and\ \citenamefont {Mast}}]{Mlynczak.2016}%
  \BibitemOpen
  \bibfield  {author} {\bibinfo {author} {\bibfnamefont {M.~G.}\ \bibnamefont
  {Mlynczak}}, \bibinfo {author} {\bibfnamefont {T.~S.}\ \bibnamefont
  {Daniels}}, \bibinfo {author} {\bibfnamefont {D.~P.}\ \bibnamefont {Kratz}},
  \bibinfo {author} {\bibfnamefont {D.~R.}\ \bibnamefont {Feldman}}, \bibinfo
  {author} {\bibfnamefont {W.~D.}\ \bibnamefont {Collins}}, \bibinfo {author}
  {\bibfnamefont {E.~J.}\ \bibnamefont {Mlawer}}, \bibinfo {author}
  {\bibfnamefont {M.~J.}\ \bibnamefont {Alvarado}}, \bibinfo {author}
  {\bibfnamefont {J.~E.}\ \bibnamefont {Lawler}}, \bibinfo {author}
  {\bibfnamefont {L.~W.}\ \bibnamefont {Anderson}}, \bibinfo {author}
  {\bibfnamefont {D.~W.}\ \bibnamefont {Fahey}}, \bibinfo {author}
  {\bibfnamefont {L.~A.}\ \bibnamefont {Hunt}},\ and\ \bibinfo {author}
  {\bibfnamefont {J.~C.}\ \bibnamefont {Mast}},\ }\bibfield  {title} {\bibinfo
  {title} {The spectroscopic foundation of radiative forcing of climate by
  carbon dioxide},\ }\href
  {https://doi.org/https://doi.org/10.1002/2016GL068837} {\bibfield  {journal}
  {\bibinfo  {journal} {Geophysical Research Letters}\ }\textbf {\bibinfo
  {volume} {43}},\ \bibinfo {pages} {5318} (\bibinfo {year} {2016})},\ \Eprint
  {https://arxiv.org/abs/https://agupubs.onlinelibrary.wiley.com/doi/pdf/10.1002/2016GL068837}
  {https://agupubs.onlinelibrary.wiley.com/doi/pdf/10.1002/2016GL068837}
  \BibitemShut {NoStop}%
\bibitem [{Note1()}]{Note1}%
  \BibitemOpen
  \bibinfo {note} {See the Supplemental Materials for the (Google Sheets) model
  at URL https://tinyurl.com/ydsdwr5w}\BibitemShut {NoStop}%
\bibitem [{\citenamefont {Reershemius}\ \emph {et~al.}(2023)\citenamefont
  {Reershemius}, \citenamefont {Kelland}, \citenamefont {Jordan}, \citenamefont
  {Davis}, \citenamefont {D'Ascanio}, \citenamefont {Kalderon-Asael},
  \citenamefont {Asael}, \citenamefont {Epihov}, \citenamefont {Beerling},
  \citenamefont {Reinhard},\ and\ \citenamefont
  {Planavsky}}]{Reershemius.2023}%
  \BibitemOpen
  \bibfield  {author} {\bibinfo {author} {\bibfnamefont {T.}~\bibnamefont
  {Reershemius}}, \bibinfo {author} {\bibfnamefont {M.~E.}\ \bibnamefont
  {Kelland}}, \bibinfo {author} {\bibfnamefont {J.~S.}\ \bibnamefont {Jordan}},
  \bibinfo {author} {\bibfnamefont {I.~R.}\ \bibnamefont {Davis}}, \bibinfo
  {author} {\bibfnamefont {R.}~\bibnamefont {D'Ascanio}}, \bibinfo {author}
  {\bibfnamefont {B.}~\bibnamefont {Kalderon-Asael}}, \bibinfo {author}
  {\bibfnamefont {D.}~\bibnamefont {Asael}}, \bibinfo {author} {\bibfnamefont
  {D.~Z.}\ \bibnamefont {Epihov}}, \bibinfo {author} {\bibfnamefont {D.~J.}\
  \bibnamefont {Beerling}}, \bibinfo {author} {\bibfnamefont {C.~T.}\
  \bibnamefont {Reinhard}},\ and\ \bibinfo {author} {\bibfnamefont {N.~J.}\
  \bibnamefont {Planavsky}},\ }\bibfield  {title} {\bibinfo {title} {A new
  soil-based approach for empirical monitoring of enhanced rock weathering
  rates},\ }\bibfield  {journal} {\bibinfo  {journal} {arXiv}\ }\href
  {https://doi.org/10.48550/arxiv.2302.05004} {10.48550/arxiv.2302.05004}
  (\bibinfo {year} {2023})\BibitemShut {NoStop}%
\bibitem [{\citenamefont {Derry}\ \emph {et~al.}(2024)\citenamefont {Derry},
  \citenamefont {Chadwick},\ and\ \citenamefont {Porder}}]{Derry2024}%
  \BibitemOpen
  \bibfield  {author} {\bibinfo {author} {\bibfnamefont {L.}~\bibnamefont
  {Derry}}, \bibinfo {author} {\bibfnamefont {O.}~\bibnamefont {Chadwick}},\
  and\ \bibinfo {author} {\bibfnamefont {S.}~\bibnamefont {Porder}},\
  }\bibfield  {title} {\bibinfo {title} {Assessment of co2 consumption from
  basalt amendments to soils},\ }\href
  {https://conf.goldschmidt.info/goldschmidt/2024/meetingapp.cgi/Paper/24194}
  {\ \bibinfo {series} {Goldschmidt} (\bibinfo {year} {2024})}\BibitemShut
  {NoStop}%
\bibitem [{\citenamefont {Mooshammer}\ \emph {et~al.}(2024)\citenamefont
  {Mooshammer}, \citenamefont {Shrieves}, \citenamefont {Chang}, \citenamefont
  {Marklein}, \citenamefont {Woodill},\ and\ \citenamefont
  {Wolf}}]{Mooshammer2024}%
  \BibitemOpen
  \bibfield  {author} {\bibinfo {author} {\bibfnamefont {M.}~\bibnamefont
  {Mooshammer}}, \bibinfo {author} {\bibfnamefont {C.}~\bibnamefont
  {Shrieves}}, \bibinfo {author} {\bibfnamefont {E.}~\bibnamefont {Chang}},
  \bibinfo {author} {\bibfnamefont {A.}~\bibnamefont {Marklein}}, \bibinfo
  {author} {\bibfnamefont {A.~J.}\ \bibnamefont {Woodill}},\ and\ \bibinfo
  {author} {\bibfnamefont {A.}~\bibnamefont {Wolf}},\ }\bibfield  {title}
  {\bibinfo {title} {A direct-measurement framework for enhanced rock
  weathering}\ }\href {https://doi.org/10.70212/cdrxiv.2024309}
  {10.70212/cdrxiv.2024309} (\bibinfo {year} {2024})\BibitemShut {NoStop}%
\end{thebibliography}

%

\end{document}